\documentstyle[prl,aps,multicol,epsf]{revtex}
\begin{document}

\draft
\title{Spin-dependent transport in a clean one-dimensional channel}
\author{C.-T.~Liang, M.~Y.~Simmons, C.~G.~Smith, G.~H.~Kim, D.~A.~Ritchie and M.~Pepper
}

\address{
Cavendish Laboratory,
Madingley Road,
Cambridge CB3 0HE,
United Kingdom
}

\date{\today}

\maketitle

\widetext
\begin{abstract}
\leftskip 54.8pt
\rightskip 54.8pt
A shoulder-like feature close to $(0.7\times 2e^{2}/h)$, 
\lq\lq the {\em 0.7 structure\/}" at zero magnetic field was observed in clean one-dimensional 
(1D) channels [K.J. Thomas {\em et al.}, Phys. Rev. Lett. 77, 135 (1996)].
To provide further understanding of this structure,
we have performed low-temperature measurements of a novel design of 1D channel with overlaying 
finger gates to study the 0.7 structure as a function of lateral confinement strength and
potential profile. We found that the structure persists when the lateral confinement strength is
changed by a factor of 2.  We have also shown that the 0.7 structure present
in two 1D channels in series behaves like a single 1D channel which shows the 0.7 structure, demonstrating 
that the 0.7 structure is not a transmission effect through a ballistic channel at zero in-plane magnetic
field.
 
\end{abstract}

\begin{multicols}{2}
\narrowtext 

Using the now well-established electrostatic squeezing split-gate technique \cite{Trevor},
it is possible to define a one-dimensional (1D) channel within a 
two-dimensional electron gas (2DEG). If the elastic 
scattering length is longer than the 1D channel length, one may observe
ballistic conductance plateaus quantized in units of $2e^2/h$ \cite{Wharam,vanWees1} at
zero magnetic field, where the factor of 2 arises from the electron spin degeneracy.
When a large magnetic field is applied parallel to the 1D channel, such that the electron
spin degeneracy is lifted, conductance plateaus quantized in units of $e^2/h$ are observed
\cite{JTN}. Quantized conductance plateaus in units of $2e^2/h$ at zero magnetic field ($e^2/h$ at 
high parallel fields) observed in a 1D channel can be well explained by cancellation of the 
Fermi velocity and 1D density of states within a single-particle picture.

In very clean 1D channels a clear plateau-like structure close to $(0.7\times 2e^{2}/h)$
has been observed at zero magnetic field $B = 0$ \cite{Thomas}. 
This {\em \lq\lq 0.7 structure"\/} whose conductance value is placed between the spin-degenerate
conductance plateau at $2e^2/h$ and the spin-split conductance plateau at $e^2/h$, cannot be
explained within a single-particle picture. The fact that the 0.7 structure evolves into a 
$(0.5\times2e^2/h)$ spin-split conductance plateau on the application of an in-plane 
magnetic field suggests the structure is related to spin.
The 0.7 structure has also been observed in 1D channels with different sample
designs \cite{Thomas,Krist,Kane}, establishing that it is a universal effect.
In particular, Kristensen {\em et al.\/}\cite{krist2} reported activated behaviour 
of the 0.7 structure as a function of temperature with a density-dependent activation 
temperature of around 2 K. Various theoretical models \cite{gold,gold2} involving partial transmission
\cite{Wang1,Wang2} and hybridization of different spin states \cite{schmeltzer} have
been proposed to explain this undisputed result, however, its exact physical
origin is still unknown. It is the purpose of this paper to report
further experimental studies.

We have designed a novel 
1D channel with three separate and independently contacted overlaying
finger gates. It is well known that the potential profile of a ballistic 1D
channel is often best described by a saddle point potential \cite{buttiker} with both lateral
and longitudinal confinement. By changing the applied voltages on the overlaying 
gate fingers above the 1D channel, we are able to
vary both the lateral confinement strength and the potential profile within
the channel. We find that the 0.7 structure is an intrinsic 
property of a clean 1D channel well over the range investigated.
Moreover, we shall present experimental evidence 
that the 0.7 structure is not a transmission effect at zero in-plane magnetic field.

Figure~1~(a) shows a schematic diagram of the device configuration. The device
was defined by electron beam lithography on the surface of the sample T258, 157~nm above a 
two-dimensional electron gas (2DEG). There is a 30~nm-thick layer of
polymethylmethacrylate (PMMA) which has been highly dosed by an electron beam,
to act as a dielectric \cite{Zailer} between the split-gate and three gate
fingers so that all gates can be independently controlled \cite{Liang}.
The carrier concentration of the 2DEG was
$1.9\times 10^{15}$ m$^{-2}$ with a mobility of 250 m$^{2}$/Vs after 
brief illumination with a red light emitting diode. The transport mean free
path is 16.5~$\mu$m, much longer than the effective 1D channel length.
Experiments were performed in a pumped $^{3}$He 
cryostat and the two-terminal conductance 
$G=dI/dV$ was measured using an ac excitation voltage of 10 $\mu$V
with standard phase-sensitive techniques.
In all cases, a zero-split-gate-voltage series resistance ($\approx$
900~$\Omega$) is subtracted from the raw data. The in-plane magnetic field 
$B_{\parallel}$ is applied parallel to the source-drain current. 
Three different samples on five different runs showed similar behaviour, and
the data that we present here are obtained from two devices A and B at three different 
cooldowns.

To demonstrate the high-quality of our 1D channel, figure~1~(b) shows the 
conductance measurements $G(V_{SG})$ as a function
of split-gate voltage $V_{SG}$ when all finger gate voltages $V_{F1}$, 
$V_{F2}$ and $V_{F3}$ are zero at $T= 300$~mK. We observe conductance plateaus at 
multiples of $2e^2/h$, with no resonant feature superimposed on top, demonstrating that we 
have a clean 1D channel in our system in which impurity scattering is negligible.
In addition, we also observe the 0.7 structure.

\begin{figure}[!t]
\begin{center}
\epsfxsize=40truemm  
\centerline{\epsffile{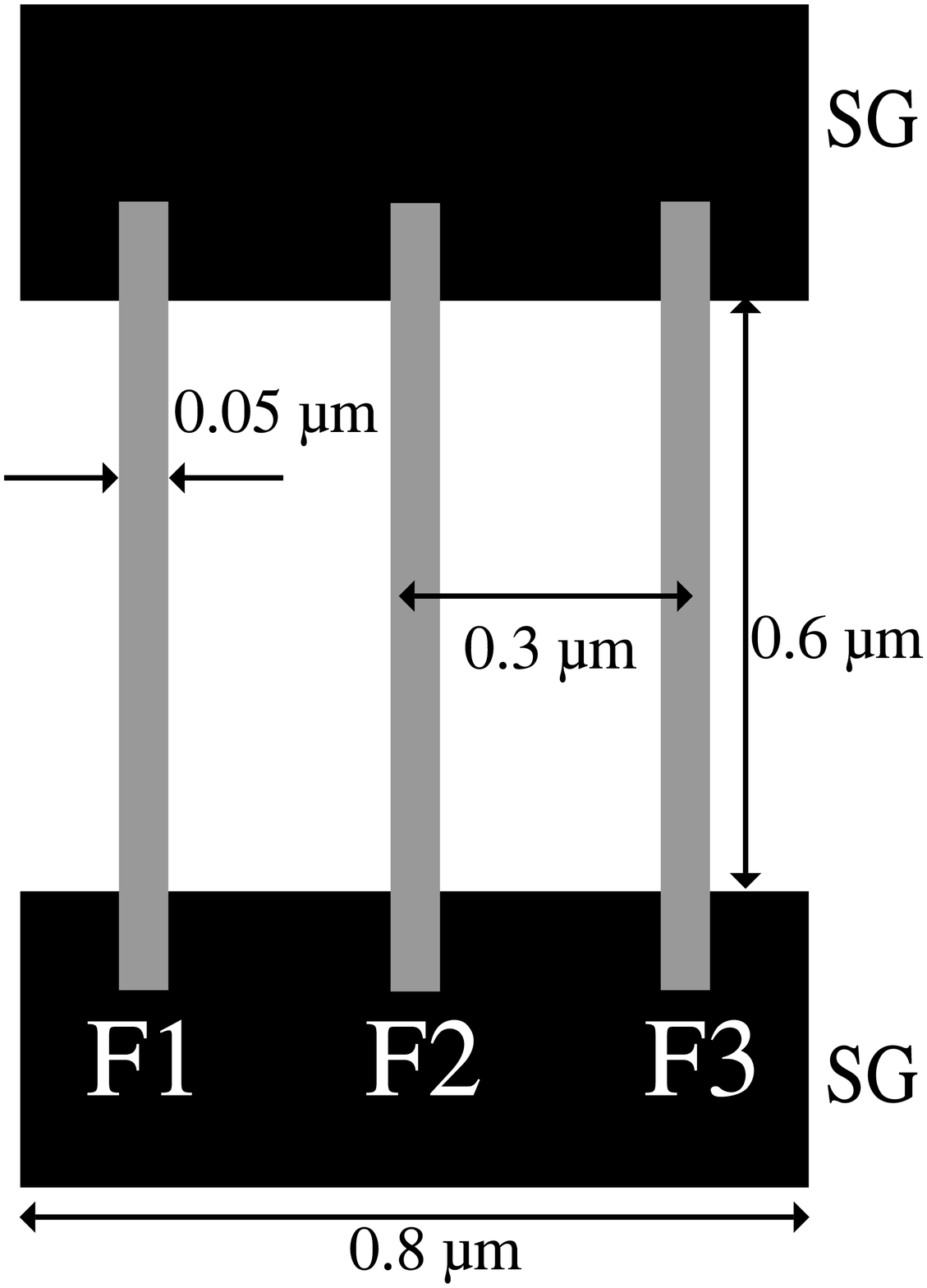}} 
\epsfxsize=80truemm  
\centerline{\epsffile{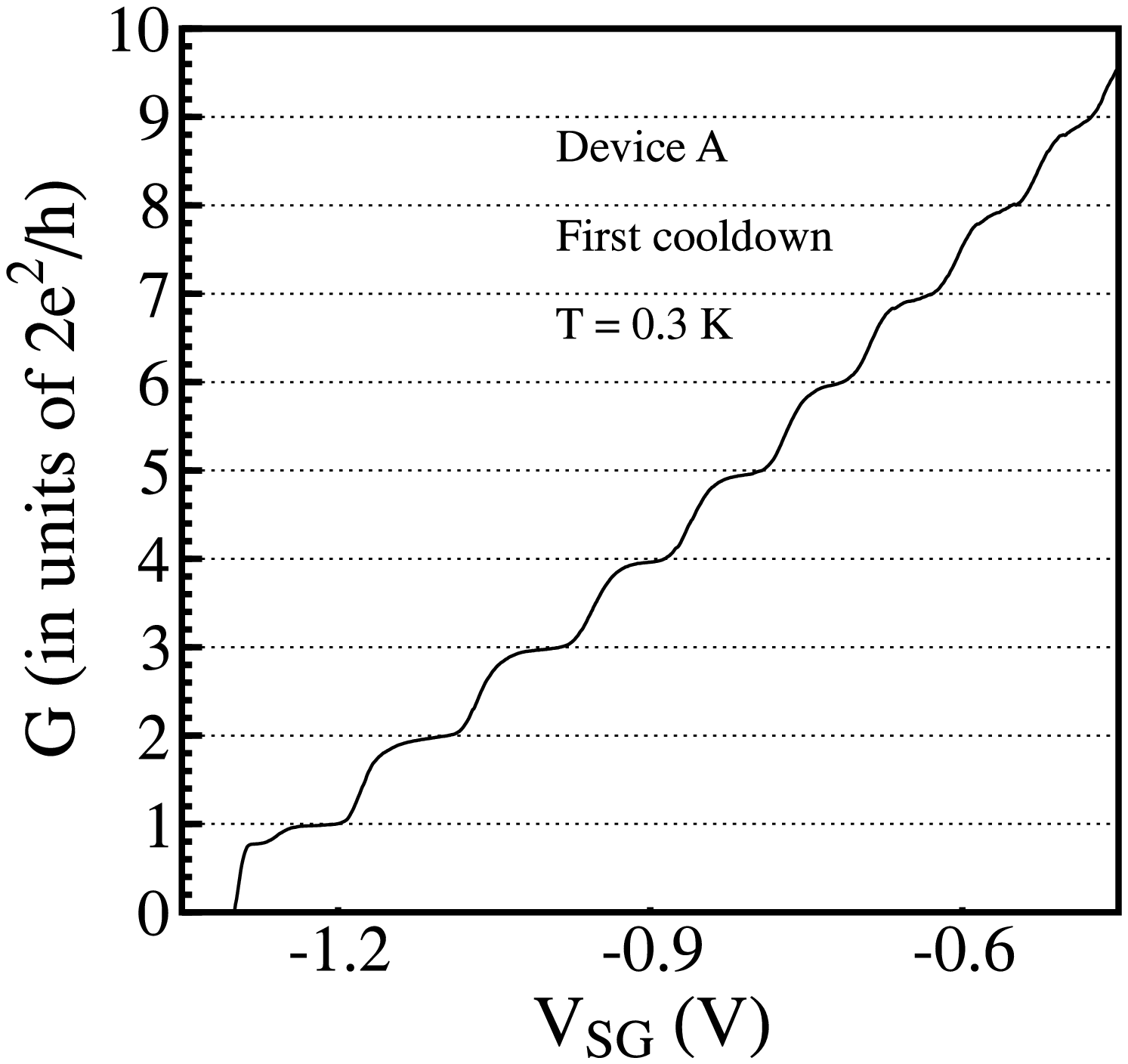}} 
\end{center}
\caption{(a) Schematic diagram showing the device configuration.
The grey regions correspond to finger gates, labelled as F1, F2, and F3 lying above 
the split-gate (labelled as SG), with an insulating layer of crosslinked PMMA in between.
(b) $G(V_{SG})$ for all finger gates at 0~V. The measurement
temperature was 300~mK.}
\label{f:1}
\end{figure}

We now describe the effects of applying a negative finger gate voltage $V_{F2}$.
Figure~2~(a) shows $G(V_{SG})$ for various voltages on
F2 while F1 and F3 are at 0~V. The results presented here
are taken at a measurement temperature of 1.2~K since the 0.7 structure
is known to be more pronounced at higher temperatures \cite{Thomas}. 
Increasing the negative voltage on F2 decreases the electron density underneath the finger 
gate. We use a technique developed by Patel {\it et al.} 
\cite{Patel} to measure the energy separation of 1D subbands from the effects of an applied dc
source-drain voltage $V_{sd}$ at various $V_{F2}$. 
The results are presented in Fig.~2~(b), and demonstrate a good linear fit $\Delta E_{1,2} = 
(0.915V_{F2}$/V+$2.71$)~(meV). It can be seen that as $V_{F2}$ is made more negative, 
the energy spacing between the first an $\Delta E_{1,2}(V_{SG})$
decreases, giving rise to the reduction in flatness of the conductance plateaus 
presented in Fig.~2~(a). Using the saddle point model \cite{buttiker}, we estimate the value
$\omega_{y}/\omega_{x}$ to decrease from 1.1 to 0.6 over the measurement range $-0.3$~V~$\geq V_{F2} 
\geq -1.8$~V. Whilst the 1D ballistic conductance
plateaus are no longer distinguishable, the shoulder-like structure close at $G=
(0.7\times2e^2/h)$ is observed to persist despite this change in the lateral confinement strength.
The data shown in figure~2~(a) provide compelling evidence 
that the 0.7 structure is intrinsic to a clean 1D channel and persists over a
wide range of lateral confinement strengths. 

\begin{figure}[!t]
\begin{center}
\epsfxsize=80truemm  
\centerline{\epsffile{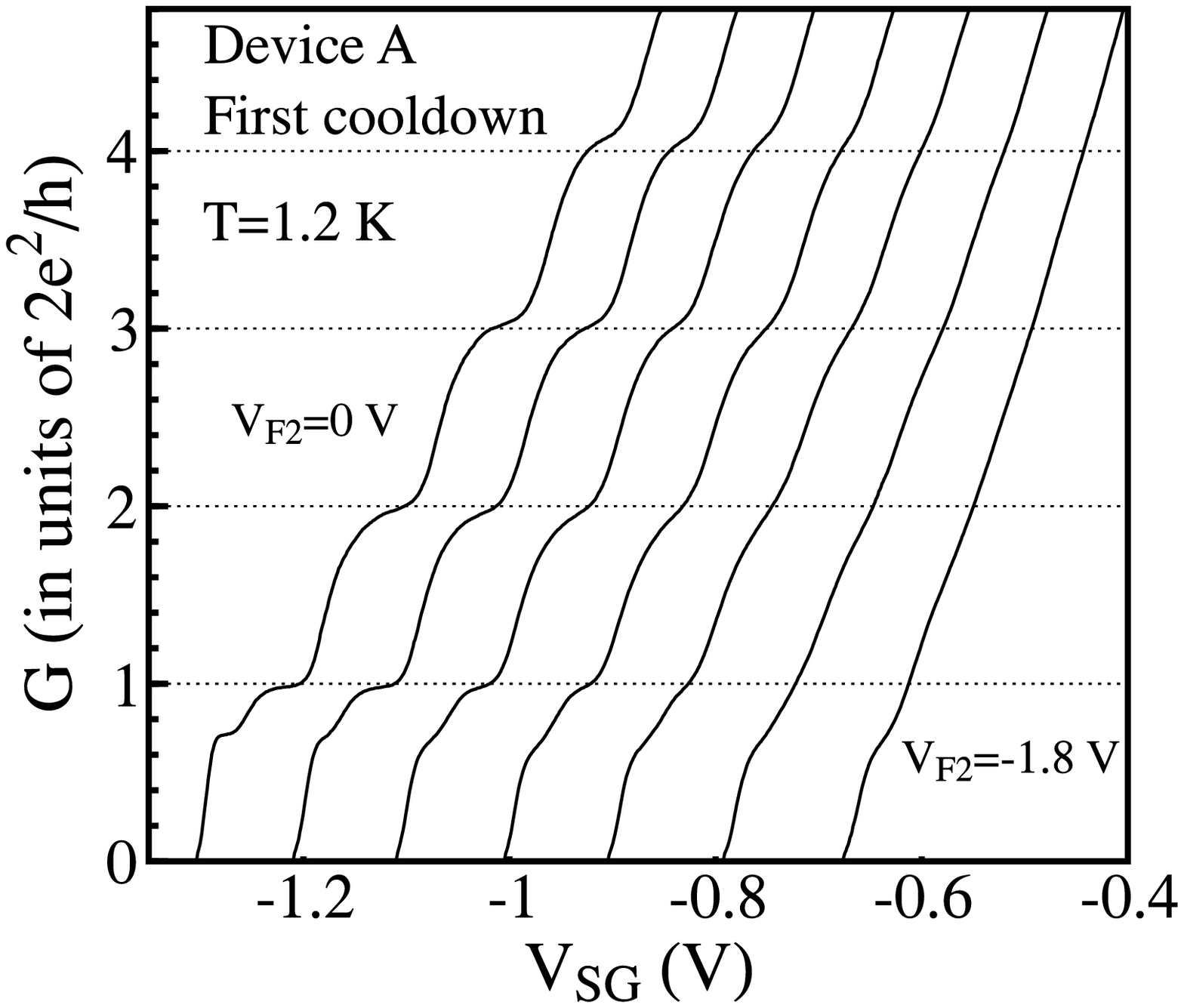}} 
\epsfxsize=50truemm  
\centerline{\epsffile{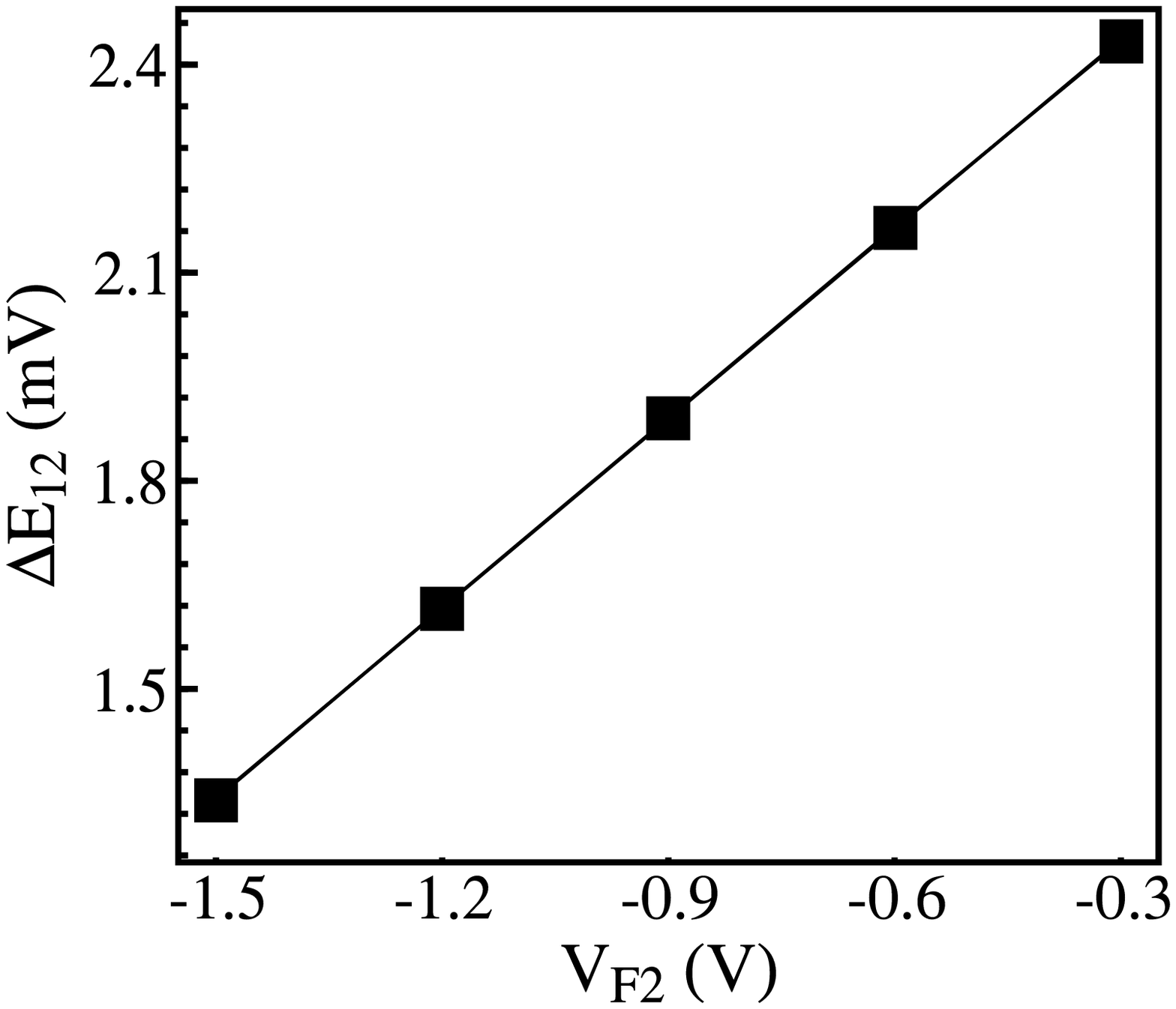}} 
\end{center}
\caption{(a) $G(V_{SG})$ for $V_{F2} = 0$ to $-1.8$~V in $0.3$~V steps when
$V_{F1} = V_{F3} = 0 $~V. The measurement temperature was 1.2~K.
(b) $\Delta E_{1,2}(V_{SG})$ (marked by
squares) determined by the source-drain bias technique. The linear fit is discussed
in the text.}
\label{f:2}
\end{figure}

To demonstrate the observed shoulder-like 
structure close to $(0.7\times2e^2/h)$, where the conductance steps are not well-quantized 
and pronounced, has the same physical origin as those observed which
coexist with well-quantized conductance steps \cite{Thomas}, we have
measured $G(V_{SG})$ at various $B_{\parallel}$. As the applied
$B_{\parallel}$ is increased, the shoulder-like feature indeed evolves
into a spin-split $(0.5\times2e^2/h)$ conductance plateau, as clearly shown in figure~3,
in agreement with early studies of Thomas {\em et al.\/}\cite{Thomas}. The fact 
that the structure at $(0.7\times2e^2/h)$ is not replicated at $0.7\times e^2/h$ when the
spin degeneracy is removed at high $B_{\parallel}$ previously reported by Thomas {\em et al.\/}
\cite{Thomas} and also shown here is evidence that the 0.7 structure is not a transmission 
effect.  

\begin{figure}[!t]
\begin{center}
\epsfxsize=70truemm  
\centerline{\epsffile{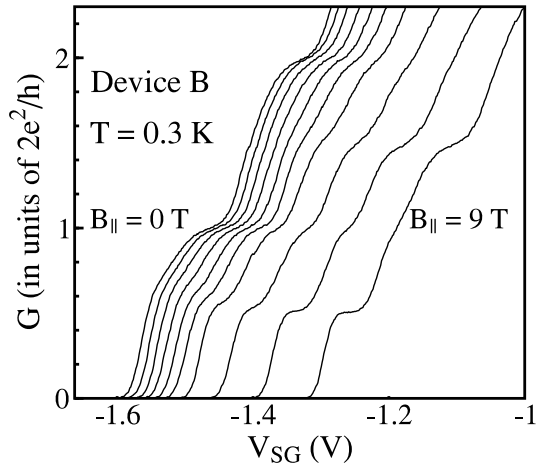}} 
\end{center}
\caption{$G(V_{SG})$ at various applied in-plane magnetic fields $B_{\parallel}$
for $V_{F2} = -1.4$~V and $V_{F1} = V_{F3} = 0 $~V. From left to right:
$B_{\parallel} = 0$ to $9$~T in $1$~T steps. Successive traces have been 
horizontally offset by 3~mV for clarity. The measurement
temperature was 300~mK.}
\label{f:3}
\end{figure}

Finally we present clear experimental evidence that the 0.7 structure  
is not a transmission effect in zero in-plane magnetic field. The solid line in figure~4 
shows $G(V_{SG})$ when $V_{F1} = -0.22$~V, $V_{F2} = 0 $~V and $V_{F3}=0$, and 
the dotted line shows $G(V_{SG})$ when $V_{F1} = 0 $~V, $V_{F2} = 0$~V 
and $V_{F3}=-0.3$~V. In both cases, we observe the 0.7 structure 
at the same $V_{SG}$ so that the two 
barriers underneath gate fingers are of the same heights. 
If we now set $V_{F1}=-0.22$~V, $V_{F2}=0$ and $V_{F3}=-0.3$~V, we obtain the dashed line
in figure~4.
From the data for $V_{F1}=V_{F2}=V_{F3} = 0$, we know that for $V_{SG} = -2.67$~V there are
two 1D subbands present in the ballistic channel defined by SG. As illustrated in Fig.~4, the 
1D channel pinches off at $V_{SG}=-2.67$~V for both cases when $V_{F1}=-0.22$~V and $V_{F3}=0$, and  
$V_{F1}=0$ and $V_{F3}=-0.3$~V. The distance between F1 and F3 is twice as much as
the distance between F1 (F3) and the underlying 2DEG. Also the presence of the grounded
F2 varies the flow of the electric field lines emitted from F1 and F3, which makes the
2DEG regions underneath F2 less affected by the fringing fields from F1 and F3.
All these results demonstrate that for $V_{F1}=-0.22$~V, $V_{F2}=0$ and $V_{F3}=-0.3$~V,
we have two narrower 1D constrictions underneath
F1 and F3 in series, present in the ballistic channel defined by SG, as illustrated in
Fig.~5. Here we estimate the constriction width underneath F1 (F3). 
Assume that the lateral (the y component) confining potential in the 1D 
channel has a form 

\begin{equation}
U (y) =U(0) + \frac{1}{2} m^{*} \omega_{y}^{2} y^{2}, 
\end{equation}
                                                      
where $m^{*} = 0.067m_{e}$ and $m_{e}$ is the electron mass.

\begin{figure}[!t]
\begin{center}
\epsfxsize=80truemm  
\centerline{\epsffile{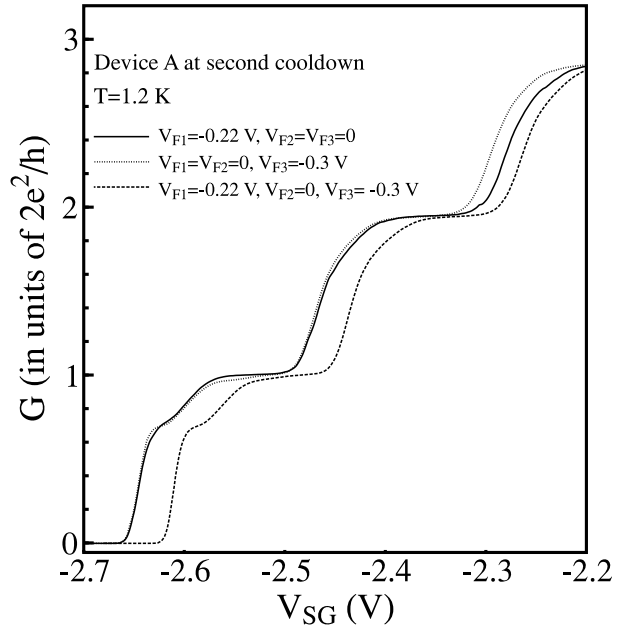}} 
\end{center}
\caption{The solid line shows $G(V_{SG})$ when $V_{F1} = -0.22$~V, $V_{F2} = 0 $~V 
and $V_{F3}=0$, and the dotted line shows $G(V_{SG})$ when $V_{F1} = 0 $~V, 
$V_{F2} = 0$ and $V_{F3}=-0.3$~V. The dashed line shows 
$G(V_{SG})$ when $V_{F1} = -0.22$~V, $V_{F2} = 0$~V and $V_{F3}=-0.3$~V, so
that we have {\em 0.7\/} structure present in two 1D channels in series. The measurement
temperature was 1.2~K.}
\label{f:4}
\end{figure}

From the source-drain biased measurements we know that $\Delta E_{1,2} = 2.434$~meV$= \hbar \omega_{y}$. The difference between
the first 1D subband and the conduction band edge is simply $\frac{1}{2} \hbar \omega_{y}$ in a simple harmonic
oscillator. Thus we calculate $\omega_{y}$ to be $3.698 \times 10^{12}$~s$^{-1}$ and $U(0)$ to be 5.78~meV.
The 1D channel width can be estimated when the energy of the first 1D subband crosses the Fermi 
energy in the bulk 2DEG ($E_{F} = 7$~meV). From this we calculate that the constriction width 
underneath F1 (F3) at the Fermi energy to be 43.3~nm. As shown in Fig.~4, we can see that the
0.7 structure is still observed when the two 1D constrictions are in series but at occurs 
at a slightly less negative $V_{SG}$. This is to be expected since two gate fingers are being
biased rather than one and there is a small degree of cross talk between F1 and F3. The
ratio of the reduction of pinch-off voltage to the initial pinch-off voltage is only
0.04/2.65=1.5\%. If the 0.7 structure were a transmission effect,
then when we have 0.7 structure present in two 1D channels in series, a 
shoulder like structure close to $0.7\times0.7=0.49 (2e^2/h)$ should be observed. 
Instead, the 0.7 structure persists and behaves as if it is like two ballistic resistors in 
series as first studied by Wharam {\em et al.} \cite{wham} and reproduced here. 
Thus our experimental results show that the 0.7 structure is {\em not\/} a transmission 
effect through a clean one-dimensional channel at zero in-plane magnetic field. 

\begin{figure}[!t]
\begin{center}                           
\epsfxsize=40truemm  
\centerline{\epsffile{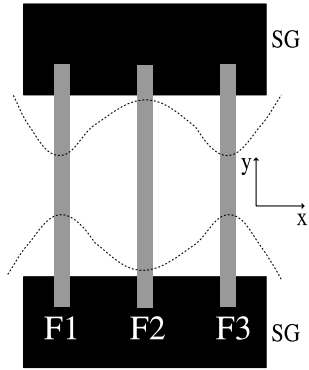}} 
\end{center}
\caption{A schematic diagram showing that applying negative voltages on F1 and 
F3 creates
two narrower 1D constrictions in series. The dotted lines indicate 
the depletion regions.}
\label{f:5}
\end{figure}

In conclusion, we have shown further supporting evidence that 0.7 structure 
is an intrinsic property of a clean one-dimensional channel, even when 
quantized ballistic plateaus are no longer distinguishable for weak lateral
confinement. Moreover, we have shown that the 
0.7 structure is {\em not\/} a transmission effect at zero in-plane magnetic field. 

This work was funded by the UK EPSRC, and in part, by the US Army Research 
Office. We thank C.H.W.~Barnes, J.T.~Nicholls, K.J.~Thomas and in particular C.J.B. Ford 
and J.H.~Davies for fruitful discussions, and J.D.F.~Franklin, J.E.F.~Frost and H.D.~Clark 
for advice and help on device fabrication at an early stage of this work. 
G.H.K. acknowledges financial support from the Skillman Fund,
Clare College. D.A.R. acknowledges support from Toshiba Research Europe Ltd.

\end{multicols}

\end{document}